# Persistent Lagrangian transport patterns in the northwestern Gulf of Mexico

Corresponding author: Matt K. Gough[1], mkgough@nps.edu

Co-authors: Francisco J. Beron-Vera[2], Maria J. Olascoaga[3], Julio Sheinbaum[1], Julien Jouanno[4], Rodrigo Duran[5]

[1]Centro de Investigacion Cientifica y de Educacion Superior de Ensenada (CICESE), Baja California, Mexico

[2]Department of Atmospheric Sciences, Rosenstiel School of Marine and Atmospheric Science, University of Miami, Miami, Florida, USA

[3]Department of Ocean Sciences, Rosenstiel School of Marine and Atmospheric Science, University of Miami, Miami, Florida, USA

[4]Laboratoire d'Etudes en Géophysique et Océanographie Spatiales (LEGOS), Toulouse, France

[5]College of Earth, Ocean and Atmospheric Sciences, Oregon State University, Corvallis, Oregon, USA

**Abstract**

Persistent Lagrangian transport patterns at the ocean surface are revealed from Lagrangian Coherent Structures (LCSs) computed from daily climatological surface current velocities in the northwestern Gulf of Mexico (NWGoM). The velocities are produced by a submesoscale permitting regional ocean model of the Gulf of Mexico. The significance of the climatological LCSs (cLCSs) is supported with ensemble-mean drifter density evolutions from simulated and historical satellite-tracked drifter trajectories. A persistent attracting barrier between the NWGoM shelf and the deep ocean is effectively identified by a hook-like pattern associated with groups of overall strongly attracting cLCSs that extend along the shelf break. Localized reductions in the attraction rate along these overall strongly attracting cLCSs proximal to cross-shore oriented cLCSs identify a pathway for potential transport across the shelf break. Groups of overall weakly-attracting cLCSs are not seen to strongly constrain material transport. Tracers originating over the shelf tend to be trapped there by the hook-like pattern as they spread cyclonically. Tracers originating beyond the shelf tend to be initially attracted to the hook-like pattern as they spread anti-cyclonically and eventually over the deep ocean. The findings have important implications for the mitigation of contaminant accidents such as oil spills.

## 1. Introduction

There are thousands of oil platforms and oil exploration sites in the northwestern Gulf of Mexico (NWGoM). Oil exploration and drilling are expected to continue, particularly since the Mexican national oil company made a major discovery in the Perdido foldbelt in 2012. The Perdido foldbelt is a geological formation that encompasses an area of nearly 40000 km$^2$ across the maritime border between the US and Mexico, lies off the continental slope at a depth of approximately 2500 m, and is rich in crude oil and natural gas. On the U.S. side of this rich oil-gas reservoir, international oil companies are already producing oil and planning expansions. Understanding surface material transport in this region is therefore critical for oil spill response and mitigation.

The surface circulation in the NWGoM is dominated by the northern extent of a persistent western boundary current that flows anti-cyclonically along the continental shelf break demarked by the 50 m isobath. The western boundary current, largely driven by wind-stress, exhibits seasonal variability with the strongest currents observed in summer and weakest currents observed in the fall (Oey 1995; Sturges 1993). Additional variability in the western boundary current is provided by interaction with eddies that have been shed off from the Loop Current and tend to propagate toward the NWGoM (Sutyrin et al. 2003; Vukovich; Crissman 1986). There is an abrupt change in circulation characteristics shoreward of the western boundary current over the NWGoM continental shelf which includes the broad Louisiana-Texas (LaTex) and relatively narrow Mexico-Texas (MexTex) shelves. Flow over the NWGoM shelf is relatively weak, generally cyclonic, largely wind-stress driven, and exhibits seasonal variability with strongest cyclonic flow during the winter and generally weaker anticyclonic flow during summer (Chu et al. 2005; Oey 1995; Ohlmann; Niiler 2005; Zavala-Hidalgo et al. 2003). Freshwater discharge from the Mississippi-Atchafalaya River system behaves as a buoyancy flow that contributes to the cyclonic flow over the shelf. A salinity front typically occurs along the outer shelf demarking the boundary between semi-freshwater discharge over the shelf and resident GoM water beyond the shelf break. During the summer the front typically resides along the LaTex shelf break and partially extends along the MexTex shelf (Zavala-Hidalgo et al. 2003). During the winter the front migrates shoreward over the LaTex shelf and extends further south along the MexTex shelf. The persistence of the front alludes to a lack of surface cross-shore exchange between the shelf and the deep ocean despite the presence of active submesoscale motions in simulations (Luo et al. 2016).

    The goal of this paper is to make use of the theory of Lagrangian Coherent Structures (LCSs) (Haller 2015), which is derived from nonlinear dynamical systems techniques, to extract robust Lagrangian transport trends from a long (18 yr) submesoscale-permitting model simulation. The LCSs are key material lines (i.e., transport barriers) that shape global transport and mixing by exerting the strongest influence on neighboring fluid. The challenge here is how to meaningfully process the information contained in such a large dataset. Lagrangian trajectory integrations require repetitive interpolation of the flow data which can be computationally expensive. This cost is compounded when these computations are repeated to illuminate general trends in Lagrangian transport. LCS studies therefore typically focus on case-studies or scenarios which verify observations with LCS maps (e.g., comparisons of LCS maps to satellite imagery of chlorophyll concentrations, sea surface temperature, or oil spill observations (Gough et al. 2016; Olascoaga; Haller 2012; Olascoaga et al. 2013)). Recently,

climatological LCSs, or cLCSs, computed from a numerical model simulation have been shown to effectively identify persistent surface transport patterns (Duran et al. 2017). Here we deviate from traditional LCS study practices by extracting LCSs from daily climatological surface current velocities constructed from a multi-year model simulation to compute cLCSs following Duran et al. (2017). There are two primary objectives to this study: 1) provide a guideline for interpreting the cLCSs and 2) provide maps of persistent sea surface Lagrangian transport patterns in the NWGoM. The maps can be applied as a guideline for predicting pollution transport at the ocean surface.

## 2. Methods

### 2.1. Model

This study incorporates an 18 year (1995 – 2012) numerical simulation of daily surface current velocities over the NWGoM. The simulation was performed with the Nucleus for European Modelling of the Ocean (NEMO) primitive equation modeling framework (Madec; the NEMO team 2008). The model is three-dimensional, utilizes curvilinear coordinates discretized on a C-grid, has 75 fixed vertical levels (12 levels in the upper 20 m and 24 levels in the upper 100 m), has a horizontal resolution of 1/36° (approximately 2.8 km), a temporal resolution of 150 s for the baroclinic time step, and encompasses the entire GoM between Latitudes 14 and 31°N and Longitudes 98 and 78°W. Surface current velocities were taken from the top level (1 m depth). Forcing at the lateral boundaries was provided by the daily MERCATOR global ocean reanalysis GLORYS2V3. Additionally, the model was forced at the surface with the Drakkar

Forcing Sets version 5.2 (DFS5.2) (Dussin et al. 2014). DFS5.2 consists of 3 hour fields of wind, temperature, and humidity, and daily fields of radiation (longwave and shortwave) and precipitation. Freshwater discharge incorporated by the model was based on monthly runoff climatology (Dai; Trenberth 2002). The simulation was initialized using geostrophic balance and the temperature and salinity fields obtained from 1/4° outputs of MERCATOR GLORYS2V3 reanalysis at day 1 January 1993 and then integrated over a 20 year period (1993 – 2012). The first two years were neglected to allow for spin-up. The original data, at 150 s, were averaged to generate daily outputs. This regional model configuration has been used in Jouanno et al. (2016) to study the formation of short-period Loop Current Frontal Eddies. We refer the reader to this publication for further details on the numerical configuration and a validation of the background state (currents and eddy kinetic energy) as well as detailed comparison of the eddy properties in the Loop Current area with in-situ and satellite observations. The model allows a partial range of submesoscale processes since these processes occur at scales $O$ (100 m) to $O$ (10 km). Because the surface mixed layer depth varies seasonally, there is a seasonal bias in the model's ability to resolve submesoscale processes.

## 2.2. Lagrangian Coherent Structures

The starting point for LCS analysis is the motion equation for fluid particles,

$$\frac{dx}{dt} = v(x,t) \qquad (1)$$

where $v(x,t)$ is the velocity of the fluid, with x and t denoting position and time. Solving this equation for fluid particles at positions $x_0$ at $t_0$, one obtains a map that takes these positions to positions $x$ at another $t$, namely,

$$F_{t_0}^t(x_0) := x(t; x_0, t_0). \tag{2}$$

A fundamental observer-independent (or objective) measure of fluid deformation is given by the Cauchy-Green strain tensor,

$$C_{t_0}^t(x_0) := \left[DF_{t_0}^t(x_0)\right]^\top DF_{t_0}^t(x_0). \tag{3}$$

The eigenvalues and eigenvectors of (3) satisfy $0 < \lambda_1(x_0) \leq \lambda_2(x_0)$ and $\xi_1(x_0) \perp \xi_2(x_0)$.

The geodesic theory of LCSs (Beron-Vera et al. 2015; Farazmand et al. 2014; Haller 2015; Haller; Beron-Vera 2012) then states that a hyperbolic LCS at time $t_0$ that attracts neighboring fluid from $t_0 - T$ through $t_0$, $T > 0$ follows as a Cauchy-Green squeezeline, namely, a parametric curve $r(s)$ such that 1)

$$\frac{dr}{ds} = \xi_1(r) \tag{4}$$

and 2) along which

$$\sqrt{\lambda_2(r)} > 1, \tag{5}$$

which measures the attraction normal to the curve. This results in a foliation of LCSs; the subset corresponding to the locally most attracting LCS is of most interest as these will have the strongest influence on neighboring fluid. The numerical computation of LCSs is well documented and a software tool is available (Onu et al. 2015). In brief, this begins by solving the trajectory equation for a grid of initial conditions spanning the region of interest and over a given time interval. The Cauchy-Green tensor is then evaluated using simple center differences over the grid of initial conditions. Cauchy-Green squeezelines are finally computed by solving (4) with the right hand side written as $sign\xi_i(r(s-\Delta)))\xi_i(r(s))$, where $\Delta$ is the integration step, which guarantees a global orientation for $\xi_i(r(s))$. Here, the width of the computational grid is set at 0.1 km. All integrations were carried out using a step-adapting fourth/fifth-order Runge-Kutta method with cubic interpolation.

## 2.3. Climatological Lagrangian Coherent Structures

The cLCS computation, as developed by Duran et al. (2017), involves performing suitable averages in two subsequent steps. In the first step we construct a daily climatological flow. The flow climatology on each day is an average of NEMO surface velocities over the 18 year simulation. With this flow we construct a series of backward-time Cauchy-Green tensor fields on sliding 7-day time windows. This sets the coherence timescale of the LCS analysis. It can be justified on the basis that a few days to a week is a critical timescale for oil-spill response which largely motivates this study. In the second step, the Cauchy-Green tensor is averaged over each month of the year-long record. The cLCS finally follow as squeezelines of these monthly averaged Cauchy-Green tensor fields. The cLCSs in this study are therefore the result of averaging implemented in both the climatological velocity computation and the averaged Cauchy-Green computation (Figures 1 and 2).

It must be noted that LCSs computed from daily climatological velocities are exactly invariant (material) for the Lagrangian kinematics sustained by these velocities. The averaging of the Cauchy—Green tensor fields formally destroys this invariance property. However, as we will show below, the invariance property and their influence on neighboring fluid are to a good extent preserved, not just for the climatological velocities, but also for the instantaneous velocities in an ensemble-mean sense.

## 2.4. Drifters

To assess the significance of the computed cLCSs from the NEMO simulation, independent (quasi) Lagrangian data will be employed. These data come from a large collection of satellite-tracked drifter trajectories obtained from multiple studies between 1994 and 2016. A total of 3207 drifter trajectories from several different sources were considered: Grand LAgrangian Experiment (GLAD) (Beron-Vera; LaCasce 2016; Olascoaga et al. 2013; Poje et al. 2014), LAgrangian Submesoscale ExpeRiment (LASER), National Oceanic and Atmospheric Administration (NOAA) Global Drifter Program (GDP) (Lumpkin; Pazos 2007), Surface Current Lagrangian-Drift Program (SCULP) (Ohlmann; Niiler 2005; Sturges et al. 2001), Horizon Marine Inc.'s EddyWatch® program, CICESE-PEMEX, NOAA South Florida Program (SFP), and U.S. Coast Guard during LASER. For details regarding the above drifter studies see Miron et al. (2017).

## 2.5. Ensemble drifter densities

We wish to show that the computed monthly cLCSs from the NEMO climatological velocities shape, to leading-order, the evolution of passive tracers under advection by NEMO instantaneous velocities. This is done by comparing the cLCSs with ensemble-mean trajectory evolutions. To scrutinize the influence of the cLCSs on tracer motion over a given month, we integrate trajectories starting every day on the given month along each of the 18 years of the NEMO simulation. We then lay a regular grid down on the domain of interest and count the number of tracer particles falling inside each bin of the grid. The counting is performed after *n* days of the trajectory initialization ignoring the specific year when that happened. The result is a 2-d histogram that provides a discrete representation of the density of the tracers on day *n*.

Figures 3 and 4 show tracer densities (normalized by the mean density) computed on day 7, 14, and 28 (left to right) during January, April, July, and October (top to bottom) for tracers initially inside the indicated box. Similarly, Figure 5 shows the combined tracer densities for the four chosen months. This method is also applied to the historical drifter data set (Figure 6). For the historical drifter density distributions, the trajectory is initiated at the time when a drifter enters the specified box. Thus, an ensemble of trajectories from the specified boxes are generated from which the drifter density distributions can be computed. It should be noted that the historical drifter data cannot provide the data density of repeated tracer deployments over the 18 year simulation. Densities computed from the simulation will be referred to as "tracer densities" and densities computed from the historical drifter record will be referred to as "drifter densities". Two boxes are used as the origin of deployed tracers, one over the LaTex shelf and the other over the Perdido foldbelt region. The tracers and drifters deployed over the LaTex shelf will be referred to as "LaTex" tracers (or drifters) and those deployed over the Perdido foldbelt region will be referred to as "Perdido" tracers (or drifters).

## 3. Results

NEMO predicts a persistent hook-like mesoscale pattern of strongly attracting cLCSs in the NWGoM which can be observed in the annual cLCSs (Figure 1) and monthly cLCSs (Figure 2). The base of the hook-like pattern extends northward along the MexTex shelf break demarked by the 50-m isobath and continues to follow the shelf break by curving eastward. The tip of the hook extends southward away from the LaTex shelf break in the vicinity of 94°W. The hook-like cLCS pattern clearly follows along the shoreward edge of the strong anti-cyclonic circulation predicted by NEMO beyond the shelf break (Figure 7). Tracer densities indicate that, in general, the hook-like cLCS pattern represents a boundary that prevents surface transport over the shelf from escaping to the deep ocean (Figures 3 and 5a-c) and represents an initially attracting boundary for circulation immediately offshore of the shelf break (Figures 4 and 5d-f). The resilient cLCS hook therefore represents a transport barrier that isolates the LaTex shelf and is consistent with prior LCS analyses (Olascoaga 2010). The southward pointing tip of the hook-like pattern represents a possible offshore transport route for tracers near the middle of the LaTex shelf break (Figures 3c and 4b) is consistent with previous studies (Duran et al. 2017; Martínez-López; Zavala-Hidalgo 2009; Zhang; Hetland 2012). East of the hook-like pattern, along the eastern half of the LaTex shelf break, groups of cLCSs composed of relatively short sections of strong attraction represent a region of general attraction (Figures 4c, f, l, and 5f). Tracers released on the shelf sufficiently away from the shelf break will not experience attraction by these cLCSs (Figures 3, and 5d-f). Rather they will spread westward over the shelf following the bend in the mostly weak convoluted cLCSs there. The meridional position of these cLCSs did not substantively migrate seasonally and is a persistent pattern also observed in the annual cLCSs (Figure 1).

None of the previously mentioned Lagrangian transport aspects are obvious from the inspection of monthly climatological velocities (Figure 7). Compare, for instance, Figure 2a with Figure 7a. No traces of the attracting barrier behavior associated with the hook-like cLCS can be determined from the climatology velocity for January. While this should not come as a big surprise, the valid question about the significance of the cLCSs still remains. We proceed in the sections that follow to show that the simulated cLCSs are not only meaningful for the simulated Lagrangian circulation, but are also supported by observations.

Consistent with the cLCS topology over the LaTex shelf, the majority of the ensemble-mean concentration of tracers initially inside the box over the LaTex shelf spread along the shelf toward the MexTex shelf (Figures 3 and 5a-c). As anticipated, and consistent with the convoluted topology of the cLCSs, an exception to this occurs in July when the tracers disperse evenly eastward and westward (Figure 3). These seasonal trends over the shelf are consistent with independent simulations by Zavala-Hidalgo et al. (2003). In the four months presented the majority of the tracers remain over the LaTex and MexTex shelves which is consistent with the identification of barriers to cross-shore transport by the cLCSs extending along the shelf break. One exception to this occurs on 28 January when small amounts of tracers are transported offshore (southward) guided by the southward pointing tip of the hook-like pattern (Figure 3c). Another exception occurs on 28 October when tracers extend southward along the MexTex shelf and begin to be flushed out through a gap in the cLCS hook (Figure 3l). It should be noted here that we are comparing cLCSs computed with a 7 day integration period to tracers that have been deployed for up to 28 days so a good corroboration would not be expected. However, the observed trapping of tracers over the shelf by processes related to the hook-like cLCSs for time periods of up to 28 days is remarkable considering the cLCS integration period is only 7 days.

The Perdido foldbelt is a deep water geological formation that is known to have great oil exploration potential. What would be the fate of oil in the unforeseen event of a spill? To further explore the significance of the computed cLCSs, and at the same time approximately address this question, we consider the ensemble-mean evolution of tracers initially inside a small box chosen over the Perdido foldbelt. For tracers deployed over the Perdido region, the rate of dispersion was much greater than those deployed over the LATEX shelf which is represented by the widely displaced and low tracer densities in the days after their deployment (Figures 4 and 5d-f). In all four months, tracers initially migrate northward and collect along the strongly attracting southern portion of the hook-like cLCSs on day 7 (Figure 4a, d, g, j). By day 14 the tracer densities decrease as they become more widely dispersed and there is a tendency for densities to increase along the western LATEX shelf break which is also northern portion of the hook-like cLCS (Figure 5e). By day 28 the tracers continue to be dispersed and there is a tendency for them to accumulate along the eastern LATEX shelf break (Figure 5f) where cLCSs tend to exhibit sections of increased attraction (e.g. Figure 2a). The Perdido tracers are likely dispersed by the western boundary current along with mesoscale eddies that have originally shed off the loop current and migrated westward into the NWGoM. Drifters from the Perdido region were more likely than those deployed over the LATEX shelf to leak shoreward through gaps in the hook-like cLCSs which was particularly evident in July (Figure 4g-i).

The aspects just described of the NEMO-simulated Lagrangian circulation are so robust that they hold independently of the season, as revealed in Figure 5, which shows annual cLCSs and the evolution of ensemble-mean annual tracer densities. The NEMO-simulated cLCSs are not only significant for the NEMO-simulated Lagrangian circulation but, quite remarkably, also to a large extent for the observed Lagrangian circulation. This is demonstrated in Figure 6, which exhibits similar drifter density distributions to those in Figure 5, but for densities constructed using historical satellite-tracked drifter trajectory data. There are fewer observed trajectories than simulated trajectories, so it is not possible to achieve a similar level of resolution, but still the agreement with the cLCSs are quite evident.

**4. Discussion**

The entire region along the shelf break is prone to baroclinic instabilities due to shear along the shoreward side of the western boundary current and a persistent salinity front that typically follows along the shelf break (Luo et al. 2016; Zavala-Hidalgo et al. 2003) and can be easily observed with satellite imagery. The instabilities occur at spatial scales much smaller than the scales demonstrated by the smooth cLCSs that extend northwestward along the MexTex shelf break (Figure 2). These structured cLCSs are therefore presumably capturing processes associated with straining of the ocean's surface by mesoscale circulation such as the persistent western boundary current and/or loop current eddies. It is not surprising that the cLCSs do not capture the instabilities due to the averaging used in their calculation. However, it is somewhat surprising that there is not more leakage across these cLCSs due to the instabilities. This result is consistent with earlier analyses of observed and simulated trajectories using altimetry-derived and high-resolution (1 km) model velocities indicating that the mesoscale circulation plays a leading role in shaping transport and mixing in the GoM (Beron-Vera; LaCasce 2016; Olascoaga et al. 2013).

While the cLCSs identify barriers in surface transport, gaps of relatively low attraction in the hook-like cLCS pattern suggest regions where there is the potential for surface transport to traverse the hook-like pattern. This is important because these low attraction gaps identify potential regions of exchange between the deep ocean and the shelf. In order for this cross-shore transport to be effective, there should also be cross-shore oriented cLCSs in the vicinity of the low attraction gaps to guide the cross-shore transport. Thus, low attraction gaps in the cLCSs provide the potential for cross-shore transport, and cross-shore oriented cLCSs provide the means for cross-shore transport. A persistent low attraction gap typically occurs where flow is redirected by a curve in the shelf break in the general vicinity of 96°W, 27°N which is captured in the annual mean (Figure 1) and monthly mean cLCSs for April, July, and October (Figure 2b-d). We speculate that shear may be responsible for the occurrence of this low attraction gap. Indeed, at this location NEMO predicts the western boundary current to curve eastward while also coming in close contact with westward flow over the inner shelf where there is presumably an increase in shear (Figure 7).

The eastern half of the LaTex shelf break is known for energetic submesoscale processes (Luo et al. 2016) and is conducive to baroclinic instabilities as fronts associated with freshwater discharge over the shelf are acted upon by shear at the shelf break. The convoluted cLCSs in this region are therefore likely identifying repeated small scale and/or submesoscale straining and confluence despite the averaging used to generate the cLCSs. In this case the mean flow is relatively weak which presumably allows smaller scale (including submesoscale) straining and confluence to notably contribute to the surface transport. The cumulative effect of these cLCSs is the identification of a less defined attracting barrier for surface transport which is best represented by the increased tracer densities along the eastern LaTex shelf break (Figure 5f).

Weak cLCSs did not consistently represent processes that strongly influence simulated tracer behavior although there are a few instances when groups of weak cLCSs can be corroborated with tracer densities. Groups of weak cLCSs could be the result of persistent weak

LCS occurrence or spurious strong LCS events that are captured in the averaging of their computations. The weak cLCSs over the deep ocean in the NWGoM are due to averaging in a region of high surface circulation variability due to the influence of migrating loop current eddies. Groups of relatively weak cLCSs appear to be identifying the offshore guidance of tracers between 26° and 24°N in early January (Figure 3a) and in October (Figure 3j-l). Note, however, that the eastward extension of tracer densities off the LaTex shelf at about 27°N are not identified by the weak hook-like cLCS pattern. Over the LaTex shelf weak cLCSs generally exhibited smaller scale variability. These cLCSs may be identifying random smaller scale turbulence and/or submesoscale processes that are captured in their averaging.

Submesoscale processes over the NWGoM have been shown to exhibit a primary energetic peak in the winter and a secondary energetic peak in the summer (Luo et al. 2016). Likewise, cLCSs over the LaTex shelf were more convoluted than those over the deep ocean and along the MexTex shelf break during the winter and summer (Figure 2a and c) than during the spring and fall (Figure 2b and d). This implies that the monthly cLCSs are capturing at least part of the range of submesoscale processes over the LaTex shelf. Like the identification of smaller scales over the LaTex shelf break previously mentioned, it was somewhat surprising that the monthly cLCSs were able to resolve processes at these scales due to the averaging incorporated in their computation. The seasonal agreement with Luo et al. (2016) suggests that cLCSs can identify a wide range of surface transport scales and that over the LaTex shelf, where mean flow is weak, surface transport characteristics by submesoscale processes can be identified by monthly cLCSs. Since the cLCSs are capable of identifying submesoscale processes, this strengthens the case that the smooth cLCSs along the MexTex shelf and over the deep ocean, where mean flow is strong, mesoscale processes are largely responsible for surface transport.

## 5. Summary and Conclusions

We have extracted robust Lagrangian transport patterns by computing climatological Lagrangian coherent structures, or cLCSs, from an 18-year simulation of surface current velocities in the Gulf of Mexico (GoM). The simulation was performed by an implementation of the submesoscale-permitting free-running NEMO (Nucleus for European Modelling of the Ocean). First we averaged the velocity record to construct daily climatological velocities. Second, we performed monthly and annual averages of the Cauchy-Green tensor computed in backward time over sliding windows along a climatological year to construct monthly and annual attracting cLCSs maps. Attention was placed on the northwestern GoM mainly because of the importance of understanding Lagrangian transport and pollutant dispersion in a region where deep water oil exploration is very important, and because NEMO predicts a robust hook-like cLCS pattern associated with a strong anticyclonic western boundary current. The hook-like cLCSs identified mesoscale patterns, were found to persist nearly year-round, and demonstrated the continental shelf is isolated from circulation beyond the shelf break. This isolation, and hence the significance of the cLCSs, was verified by ensemble-mean Lagrangian trajectories generated numerically by the NEMO simulation as well as observed by historical satellite-tracked surface drifting buoys. In other words, simulations and observations both indicate that the continental shelf is difficult to be reached by pollutants originating from sources beyond the shelf break, but can be heavily impacted by pollutants released over the shelf. The significance of the mesoscale cLCSs is quite surprising given the intense submesoscale activity sustained by NEMO. Further investigation is required to asses why.

**Acknowledgements**

We would like to thank Favio Medrano from the Departamento de Computación at CICESE for his help in speeding up the LCS computations. We would also like to thank Alejandro Dominguez (CICESE) for his input and conversations. The GLAD (doi:10.7266/N7VD6WC8) and LASER (doi:10.7266/N7W0940J) drifter trajectory datasets are publicly available through the Gulf of Mexico Research Initiative Information and Data Cooperative (GRIIDC) at https://data.gulfresearchinitiative.org. The NOAA/GDP dataset is available at http://www.aoml.noaa.gov/phod/dac. The drifter trajectory data from Horizon Marine Inc.'s EddyWatch® program been obtained as a part of a data exchange agreement between Horizon Marine Inc. and CICESE-Pemex. The CICESE-Pemex "Caracterización Metoceánica del Golfo de México" project was funded by PEMEX contracts SAP-428217896, 428218855, and 428229851. The quality control and post-processing of the CICESE-Pemex data were carried out by Paula Garcia and Argelia Ronquillo. Support for this work was provided by Consejo Nacional de Ciencia y Tecnología (CONACyT)—Secretaría de Energía (SENER) grant 201441 (MKG, FJBV, MJO, JS and JJ) as part of the Consorcio de Investigación del Golfo de México (CIGoM) and the Gulf of Mexico Research Initiative (FJBV and MJO) as part of the Consortium for Advanced Research of Transport of Hydrocarbons in the Environment (CARTHE). Work by RD was in support of the National Energy Technology Laboratory's ongoing research under the Offshore Field Work Proposal DOE NETL FY14-17 under the RES contract DE-FE0004000.

References

Beron-Vera, F. J., and J. H. LaCasce, 2016: Statistics of Simulated and Observed Pair Separations in the Gulf of Mexico. *Journal of Physical Oceanography*, **46,** 2183-2199.

Beron-Vera, F. J., M. J. Olascoaga, G. Haller, M. Farazmand, J. Triñanes, and Y. Wang, 2015: Dissipative inertial transport patterns near coherent Lagrangian eddies in the ocean. *Chaos: An Interdisciplinary Journal of Nonlinear Science*, **25,** 087412.

Chu, P. P., L. M. Ivanov, and O. V. Melnichenko, 2005: Fall–Winter Current Reversals on the Texas–Louisiana Continental Shelf. *Journal of Physical Oceanography*, **35,** 902-910.

Dai, A., and K. E. Trenberth, 2002: Estimates of Freshwater Discharge from Continents: Latitudinal and Seasonal Variations. *Journal of Hydrometeorology*, **3,** 660-687.

Duran, R., F. J. Beron-Vera, and M. J. Olascoaga, 2017: Quasi-steady Lagrangian transport patterns in the Gulf of Mexico. *arXiv:1704.02389v3*.

Dussin, R., B. Barnier, and L. Brodeau, 2014: The making of Drakkar forcing set DFS5  29 pp.

Farazmand, M., D. Blazevski, and G. Haller, 2014: Shearless transport barriers in unsteady two-dimensional flows and maps. *Physica D: Nonlinear Phenomena*, **278–279,** 44-57.

Gough, M. K., A. Reniers, M. J. Olascoaga, B. K. Haus, J. MacMahan, J. Paduan, and C. Halle, 2016: Lagrangian Coherent Structures in a coastal upwelling environment. *Continental Shelf Research*, **128,** 36-50.

Haller, G., 2015: Lagrangian Coherent Structures. *Annual Review of Fluid Mechanics*, **47,** 137-162.

Haller, G., and F. J. Beron-Vera, 2012: Geodesic theory of transport barriers in two-dimensional flows. *Physica D: Nonlinear Phenomena*, **241,** 1680-1702.

Jouanno, J., J. Ochoa, E. Pallàs-Sanz, J. Sheinbaum, F. Andrade-Canto, J. Candela, and J.-M. Molines, 2016: Loop Current Frontal Eddies: Formation along the Campeche Bank and Impact of Coastally Trapped Waves. *Journal of Physical Oceanography*, **46,** 3339-3363.

Lumpkin, R., and M. Pazos, 2007: Measuring surface currents with Surface Velocity Program drifters: the instrument, its data, and some recent results. *Lagrangian Analysis and Prediction of Coastal and Ocean Dynamics*, Cambridge Univ. Press, 39-67.

Luo, H., A. Bracco, Y. Cardona, and J. C. McWilliams, 2016: Submesoscale circulation in the northern Gulf of Mexico: Surface processes and the impact of the freshwater river input. *Ocean Modelling*, **101,** 68-82.

Madec, G., and the NEMO team, 2008: *NEMO ocean engine.* 27 ed., 407 pp.

Martínez-López, B., and J. Zavala-Hidalgo, 2009: Seasonal and interannual variability of cross-shelf transports of chlorophyll in the Gulf of Mexico. *Journal of Marine Systems*, **77,** 1-20.

Miron, P., F. J. Beron-Vera, M. J. Olascoaga, J. Sheinbaum, P. Pérez-Brunius, and G. Froyland, 2017: Lagrangian dynamical geography of the Gulf of Mexico. *Scientific Reports*, **7,** 7021.

Oey, L.-Y., 1995: Eddy- and wind-forced shelf circulation. *J. Geophys. Res.*, **100,** 8621-8637.

Ohlmann, J. C., and P. P. Niiler, 2005: Circulation over the continental shelf in the northern Gulf of Mexico. *Progress In Oceanography*, **64,** 45-81.

Olascoaga, M. J., 2010: Isolation on the West Florida Shelf with implications for red tides and pollutant dispersal in the Gulf of Mexico. *Nonlin. Processes Geophys.*, **17,** 685-696.

Olascoaga, M. J., and G. Haller, 2012: Forecasting sudden changes in environmental pollution patterns. *Proceedings of the National Academy of Sciences*, **109,** 4738-4743.

Olascoaga, M. J., and Coauthors, 2013: Drifter motion in the Gulf of Mexico constrained by altimetric Lagrangian coherent structures. *Geophysical Research Letters*, **40,** 6171-6175.

Onu, K., F. Huhn, and G. Haller, 2015: LCS Tool: A computational platform for Lagrangian coherent structures. *Journal of Computational Science*, **7,** 26-36.

Poje, A. C., and Coauthors, 2014: Submesoscale dispersion in the vicinity of the Deepwater Horizon spill. *Proceedings of the National Academy of Sciences*, **111,** 12693-12698.

Sturges, W., 1993: The annual cycle of the western boundary current in the Gulf of Mexico. *Journal of Geophysical Research: Oceans*, **98,** 18053-18068.

Sturges, W., P. Pearn, P. P. Niiler, and R. H. Weisberg, 2001: Northeastern Gulf of Mexico inner shelf circulation study. Final Report, MMS Cooperative Agreement 14-35-0001-30787. OCS Report MMS 2001  103 90 pp.

Sutyrin, G. G., G. D. Rowe, L. M. Rothstein, and I. Ginis, 2003: Baroclinic Eddy Interactions with Continental Slopes and Shelves. *Journal of Physical Oceanography*, **33,** 283-291.

Vukovich, F. M., and B. W. Crissman, 1986: Aspects of warm rings in the Gulf of Mexico. *Journal of Geophysical Research: Oceans*, **91,** 2645-2660.

Zavala-Hidalgo, J., S. L. Morey, and J. J. O'Brien, 2003: Seasonal circulation on the western shelf of the Gulf of Mexico using a high-resolution numerical model. *Journal of Geophysical Research: Oceans*, **108,** n/a-n/a.

Zhang, Z., and R. Hetland, 2012: A numerical study on convergence of alongshore flows over the Texas-Louisiana shelf. *Journal of Geophysical Research: Oceans*, **117,** n/a-n/a.

Figures

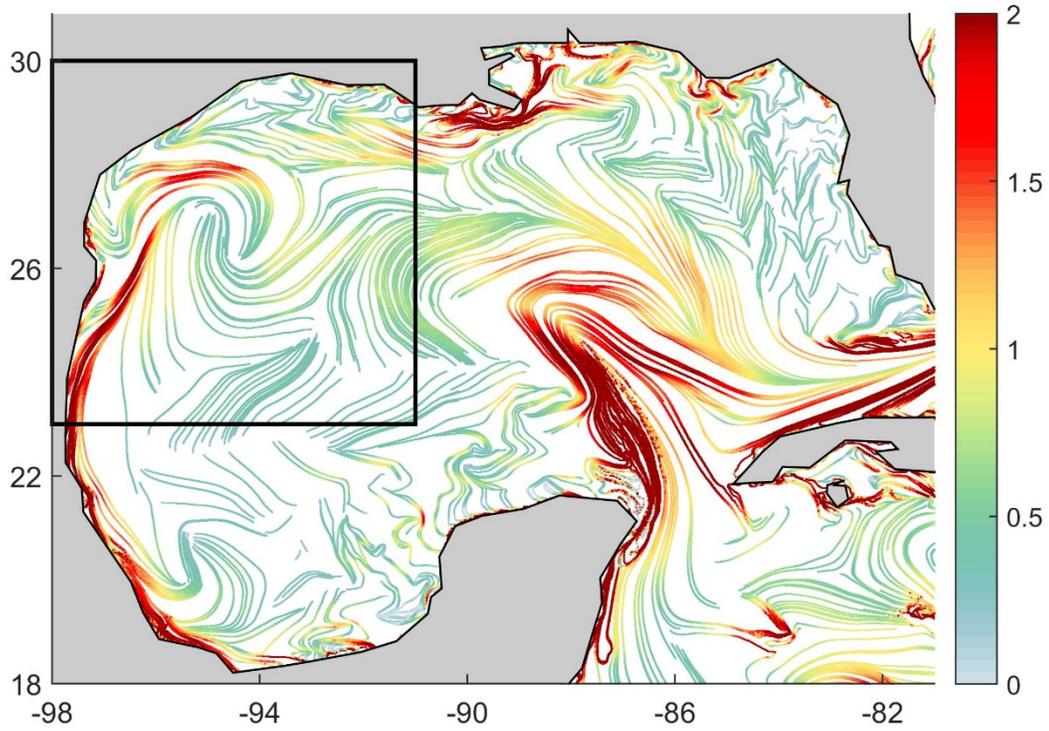

**Figure 1.** Annual cLCS map for the GoM. Colors indicate strength of attraction quantified by $\log\sqrt{\lambda_2}$. Every other squeezeline is displayed. The black box indicates NWGoM region of interest.

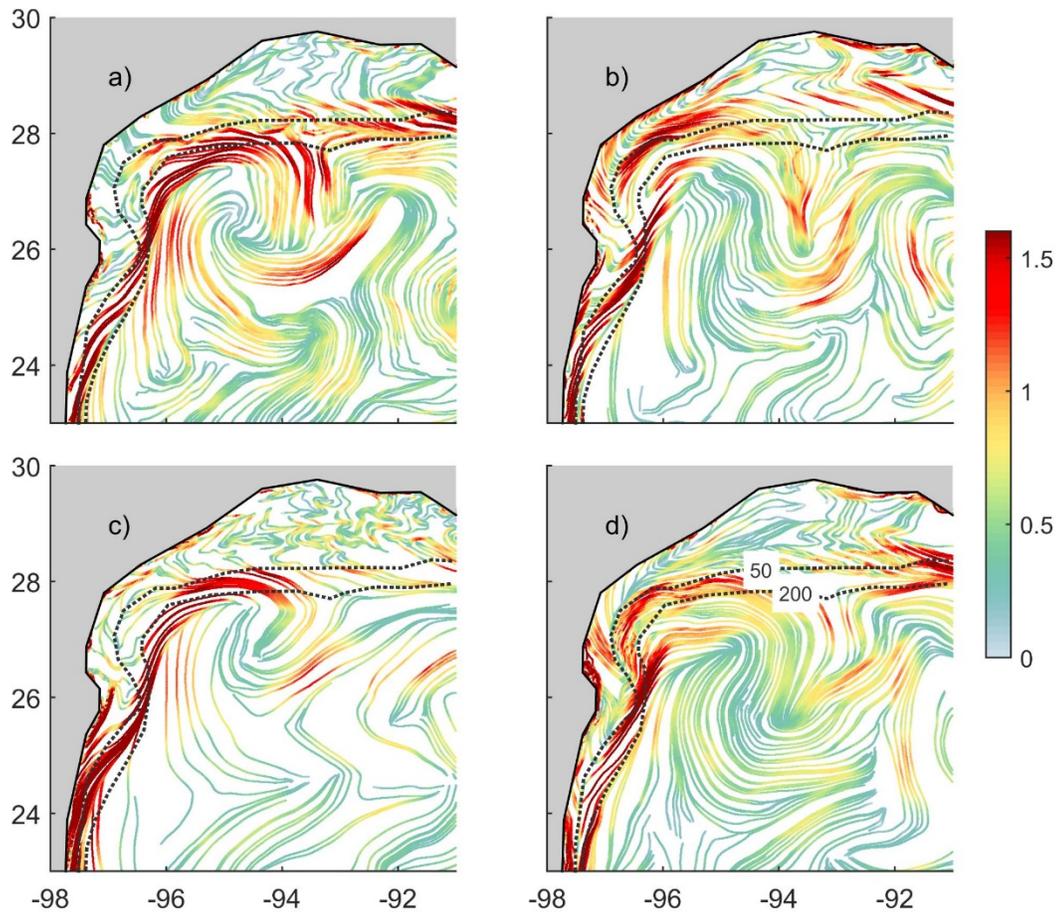

**Figure 2.** Monthly cLCSs maps for a) January, b) April, c) July, d) October. Colors indicate strength of attraction quantified by $\log\sqrt{\lambda_2}$. Every other squeezeline is displayed. The 50 and 200 m isobaths are demarked by the black dashed lines.

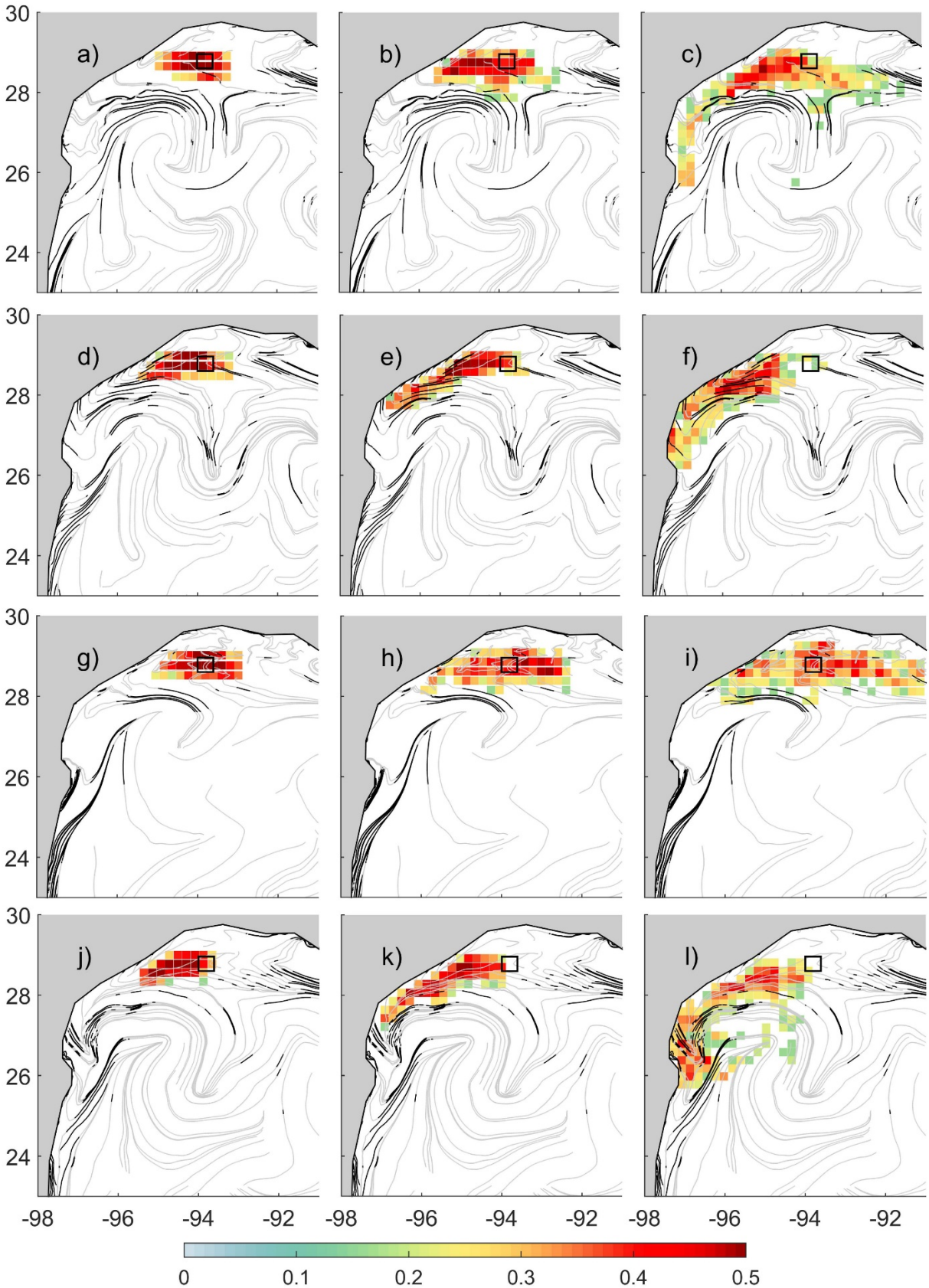

**Figure 3.** Ensemble tracer densities from simulated trajectories over the LATEX shelf for days 7, 14, and 28 (a-c) January, (d-f) April, (g-i) July, and (j-l) October. The density has been transformed to the 1/4 power (i.e. $d^{1/4}$). The underlying monthly cLCS maps correspond to figure 2.a-d. and are grey-scaled according to strength of attraction quantified by $\log\sqrt{\lambda_2}$: black (> 1.0), grey (< 1.0). Every third squeezeline is displayed. Black boxes indicate position of initial tracer deployments on the first of each month.

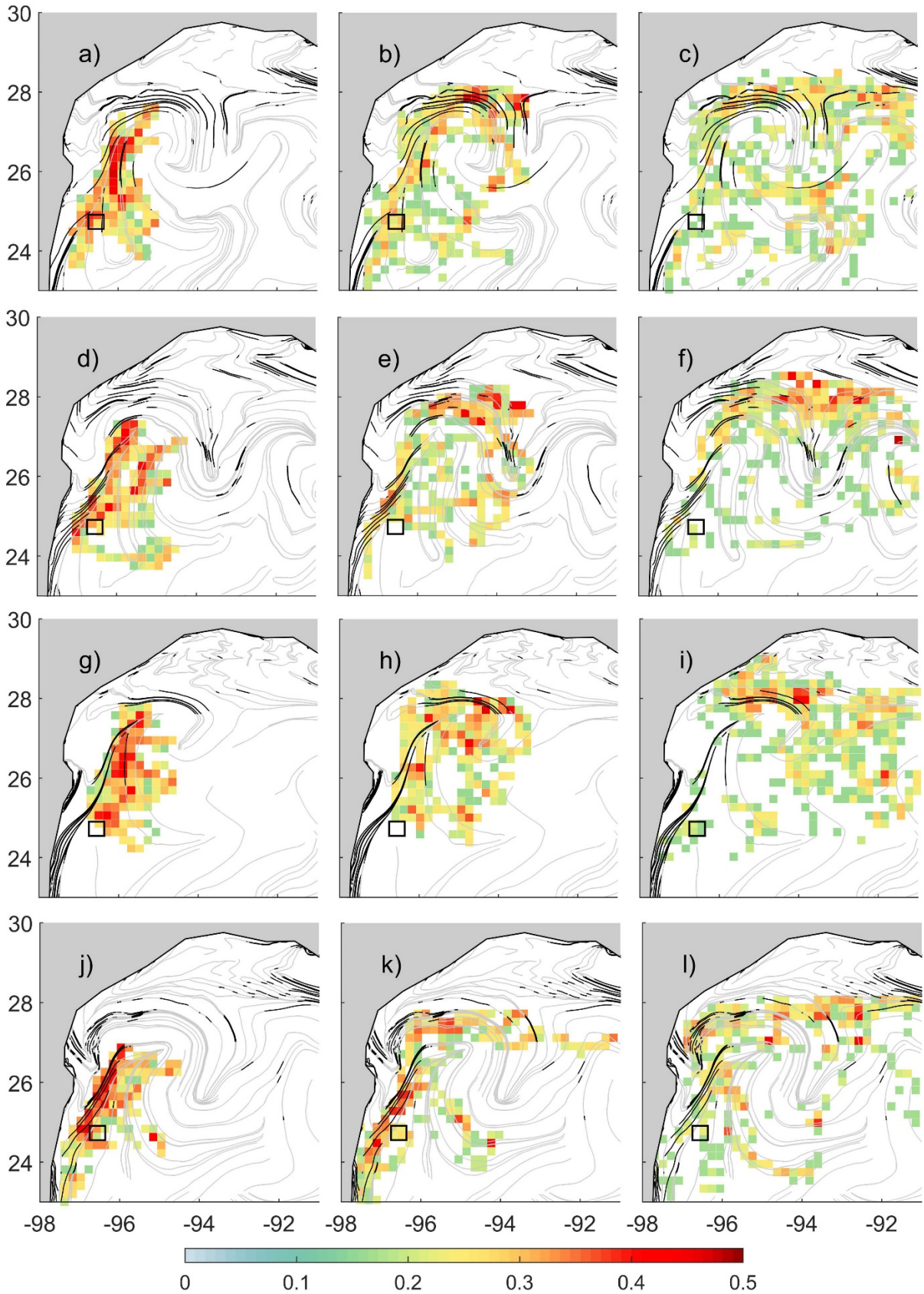

**Figure 4.** Ensemble tracer densities from simulated trajectories over the Perdido region for days 7, 14, and 28 (a-c) January, (d-f) April, (g-i) July, and (j-l) October. The density has been transformed to the 1/4 power (i.e. $d^{1/4}$ ). The underlying monthly cLCS maps correspond to figure 2.a-d. and are grey-scaled according to strength of attraction quantified by $\log\sqrt{\lambda_2}$: black (> 1.0), grey (< 1.0). Every third squeezeline is displayed. Black boxes indicate position of initial tracer deployments on the first of each month.

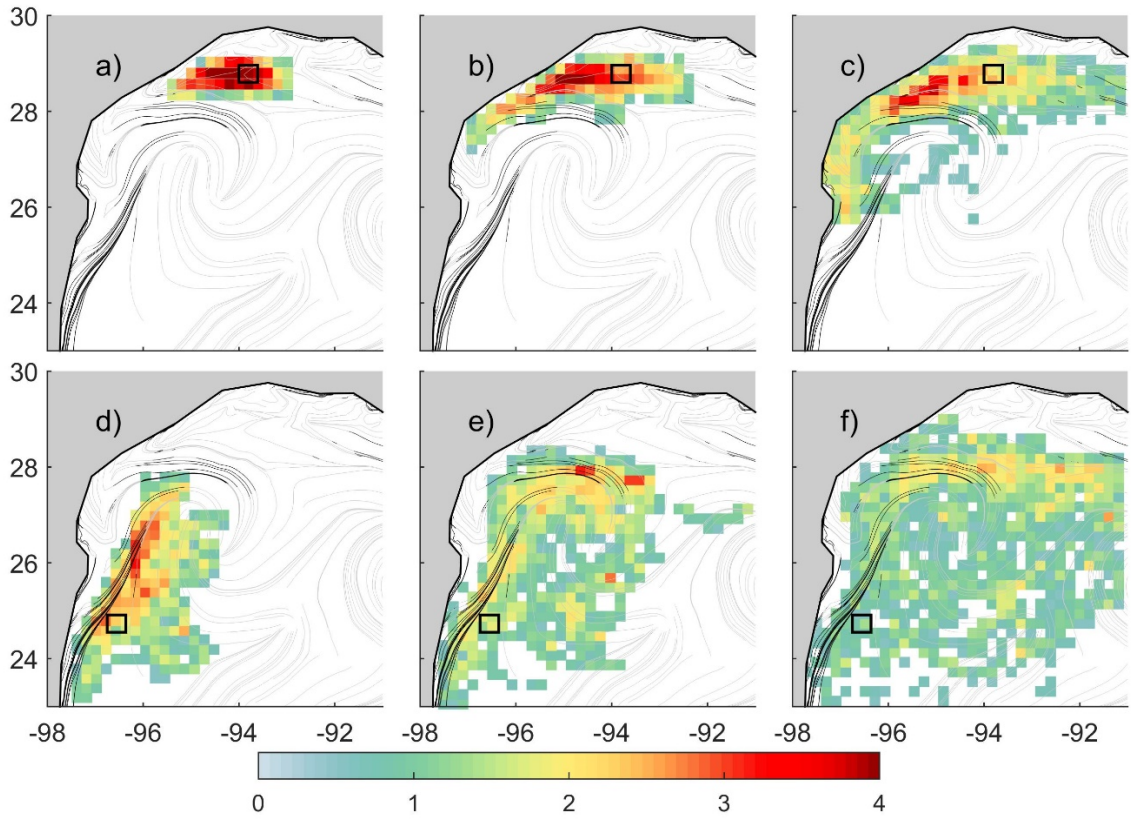

**Figure 5.** Combined ensemble tracer densities from simulated trajectories for days 7, 14, and 28 over the (a-c) LATEX shelf and (d-f) Perdido region. The density has been transformed to the 1/3 power (i.e. $d^{1/3}$). The underlying annual cLCS maps, which correspond to figure 1, are grey-scaled according to strength of attraction quantified by $\log\sqrt{\lambda_2}$: black (> 1.0), grey (< 1.0). Every third squeezeline is displayed. Black boxes indicate position of initial tracer deployments on the first of each month.

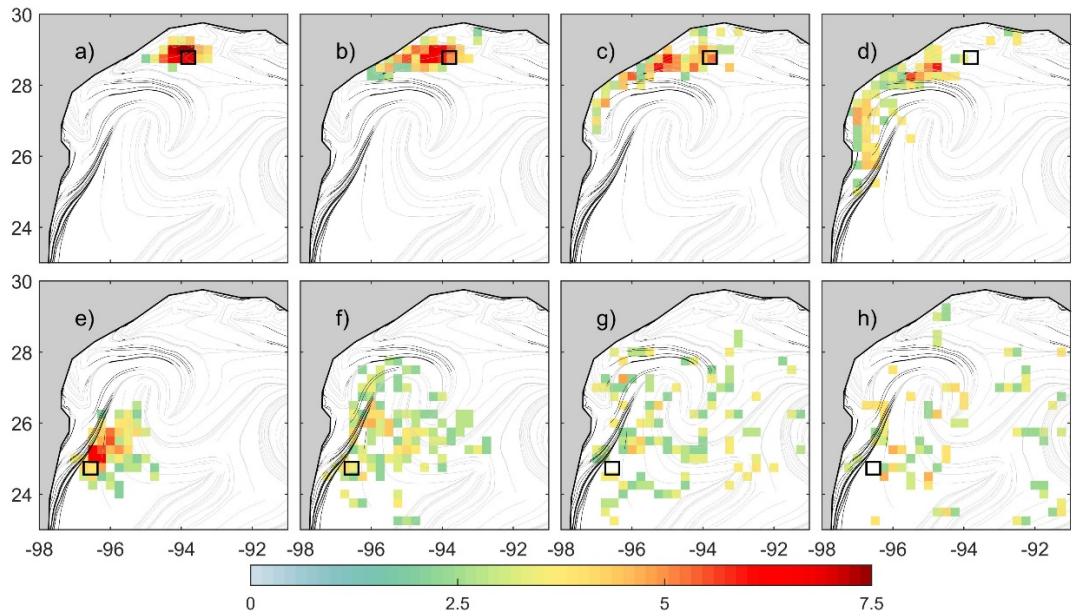

**Figure 6.** Historical drifter density distributions for days 3, 7, 14, and 28 from the time drifter trajectories passed through boxes over the (a-d) LATEX shelf and (e-h) Perdido region. The density has been transformed to the 1/3 power (i.e. $d^{1/3}$). The underlying annual cLCSs maps, which correspond to figure 1, are grey-scaled according to strength of attraction quantified by $\log\sqrt{\lambda_2}$: black (> 1.0), grey (< 1.0). Every third squeezeline is displayed.

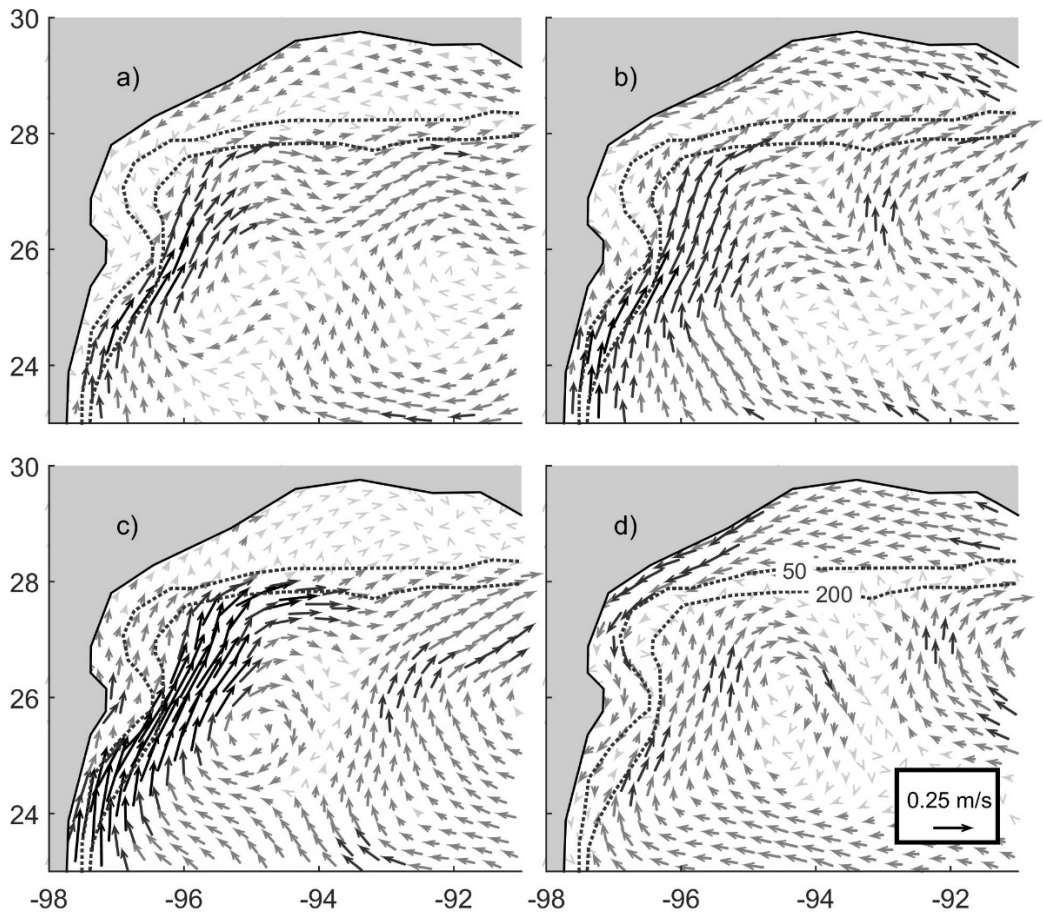

**Figure 7.** Monthly mean surface current velocity vectors for a) January, b) April, c) July, d) October. The 50 and 200 m isobaths are demarked by the black dashed lines.